\documentclass[10pt,letterpaper]{article}
\usepackage[top=0.85in,left=2.75in,footskip=0.75in]{geometry}

\usepackage{amsmath,amssymb}

\usepackage{changepage}

\usepackage[utf8x]{inputenc}

\usepackage{textcomp,marvosym}

\usepackage{cite}

\usepackage{nameref,hyperref}

\usepackage{microtype}
\DisableLigatures[f]{encoding = *, family = * }

\usepackage[table]{xcolor}

\usepackage{array}

\newcolumntype{+}{!{\vrule width 2pt}}

\newlength\savedwidth

\raggedright
\usepackage[aboveskip=1pt,labelfont=bf,labelsep=period,justification=raggedright,singlelinecheck=off]{caption}

\makeatletter
\renewcommand{\@biblabel}[1]{\quad#1.}
\makeatother

\date{}

\usepackage{lastpage,fancyhdr,graphicx}
\usepackage{epstopdf}

\begin{document}
\vspace*{0.2in}

\title{Manuscript-2}

\begin{flushleft}
{\Large
\textbf\newline{Collective Motion of Predictive Swarms} 
}
\newline
\\
Nathaniel Rupprecht \textsuperscript{1},
Dervis Can Vural \textsuperscript{1*},
\\
\bigskip
\textbf{1} Department of Physics, University of Notre Dame, South Bend, Indiana, USA
\\
\bigskip

* Corresponding author \\
Email: dvural@nd.edu (DCV)

\end{flushleft}
\section*{Abstract}
Theoretical models of populations and swarms typically start with the assumption that the motion of agents is governed by the local stimuli. However, an intelligent agent, with some understanding of the laws that govern its habitat, can anticipate the future, and make predictions to gather resources more efficiently. Here we study a specific model of this kind, where agents aim to maximize their consumption of a diffusing resource, by attempting to predict the future of a resource field and the actions of other agents. Once the agents make a prediction, they are attracted to move towards regions that have, and will have, denser resources.  We find that the further the agents attempt to see into the future, the more their attempts at prediction fail, and the less resources they consume. We also study the case where predictive agents compete against non-predictive agents and find the predictors perform better than the non-predictors only when their relative numbers are very small. We conclude that predictivity pays off either when the predictors do not see too far into the future or the number of predictors is small.


\section*{Introduction}

One of the steadfast pillars of science is the notion of causality, that the past determines the future. However, the fact that the world can be \emph{correctly} modeled leads to an interesting complication: a intelligent system can understand the laws governing its world, infer the outcomes of available choices, and decide how to behave in the present according to a predicted future.

Many-body physics has extended its applicability from equilibrium systems \cite{pathria1986statistical}  to externally driven systems \cite{zwanzig2001nonequilibrium,schuster2013nonequilibrium}, to self-driven systems \cite{tiribocchi2015active,marchetti2013hydrodynamics,ray2014casimir}. Can we extend this domain further, from describing passive, reactive and active states of matter, towards ``intelligent'' or ``predictive'' states of matter? Are there general laws and constraints govern these states?

Predictivity plays an important role in various social and biological systems. One example is a walking crowd. It has been found that walking people choose paths that will not result in a collision within the next few seconds \cite{karamouzas2014universal}. A second familiar predictive system is the stock market. Since price changes are driven by the amount of surprise created by financial news, in order to beat the returns of others, investors must predict both the news and the reaction of others to the news \cite{fostel2012does,joshi2003concepts,maslowska2012does}. It is well known in the economics literature that prediction and strategy anticipation can have complex effects and induce volatility in markets \cite{mackenzie2008engine}. Another example of predictivity is the pursuit-evasion problem. As in the former two examples, an intelligent agent, for example a prey, must base its motion not only on a predicted future, but also on the predicted future of another predictor, say, a predator. Scientific models describing social systems can couple to and modify their subjects too. Depending on the nature of the constituent agents, this can lead to either the failure or reinforcement of the model prediction \cite{vural2011models}.

The field of differential game theory aids in describing systems in which strategy and anticipation plays an important role \cite{ramachandran2012stochastic,basar1999dynamic}. There is an emerging literature on multiplayer differential games \cite{yeung2006cooperative}. However, classical differential game theory is typically concerned with a finite number of discrete actors (typically two) instead of a continuous distribution of actors or a very large number of actors. Furthermore, these games are typically formulated solely in terms of schematic payoff matrices rather than taking into account realistic physical interactions and environments.

Here we study the implications of predictivity on the properties of an active swarm. In our model, agents in the swarm attempt to predict the future, and aim to maximize their consumption of a diffusing resource by being attracted to regions of the space with high resource concentration in the present as well as the projected future. We find that attempting to predict the future makes the agents' predictions unreliable.  Furthermore, their success, as measured in terms of resource consumption, also decreases. Next, we study how a predictive population behaves when mixed with a non-predictive population that is competing for the same resource. In this case, we find that predictive agents outperform the non-predictive agents only when they constitute a minority of the total population. In this case, seeing further into the future allows the predictive agents to consume more of the resource. However, increasing the number of predictive agents in the population decreases both the prediction quality and the resource consumption of the predictors.

While our model is couched in a biological setting (e.g. describing agents consuming resources), is not intended to recreate the exact behavior of any specific biological system. Instead, we are interested in the theoretical constraints on prediction-making when predictors are coupled to the system they aim to predict \cite{vural2011models}, especially when there are multiple predictors involved.

The idea of acting in the present to maximize entropy over an entire path in the next time $\tau$ is studied in \cite{wissner2013causal}, and is qualitatively similar to what we will discuss in this work, where agents aim to maximize their resource consumption over their trajectory in the next time $\tau$. A major difference is that in \cite{wissner2013causal}, the maximized quantity is statistical path entropy, based on macrostate probabilities, whereas in our case, we will have agents attempting to maximize their consumption of a resource. No calculation of or reference to probabilities is necessary.

\section*{Materials and methods}

\subsection*{The Model} We study the properties of a population $n(\vec{r},t)$ where every agent aims to maximize its consumption of a resource $\phi(\vec{r},t)$. We assume that $\phi$ diffuses, and is absorbed by the agents according to the standard diffusion equation 
\begin{align}
\frac{d\phi}{dt} = D\nabla^2 \phi- \gamma \phi\cdot n(\vec{r}, t).
\label{resourceEvolution}
\end{align}

Secondly, we assume that every agent knows this equation, can solve it to predict the future of $\phi$, and will be attracted to regions of space where $\phi$ will be large.  The motion of the agents is governed by a nonlocal integral equation that requires the current motion of the agent to be consistent with the future that stems from its actions. For the sake of simplicity, we assume that the predictive agents move with a velocity $\vec{v}(\vec{r},t)=c\hat{u}$ of constant magnitude $c$. Thus, the predictive aspect of the agents come into play only through their selection of swimming direction, $\hat{u}(t)=\vec{M}/|M|$. The swimming direction at time $t$ is determined by the resource distribution $\phi$ between the present, $t$, and a future time $t+\tau$,
\begin{align}
\vec{M}(\vec{r},t)=\int\limits_{0}^{\tau} d t^\prime\!\!\!\int\limits_{0<|\vec{r}'-\vec{r}|<c\cdot t^\prime} \!\!\!\!\!\!\!\!d^2\!\vec{r}' \cdot \phi(\vec{r}',t+t') \hat{w}(\vec{r}'-\vec{r}) F(R)
\label{MEquation}
\end{align}
where $R=\sqrt{(\vec{r'}-\vec{r})^2+\lambda \cdot t'^2}$, and $\hat{w}(\vec{r}) = \vec{r}/|\vec{r}|$ is a unit vector and $F(R)$ is a force scaling, taken to be $1/R^2$ (we show in the Appendix that the exact form of $F(R)$ is not important when $\tau$ is ``small enough'' in the sense that the form of $F$ comes in as a correction to $\vec{M}$ at $\mathcal{O}(\tau^2)$). Throughout we use the numerical values $c=1, \lambda=1$.

In summary, agents estimate the resource distribution $\tau$ ahead, and experience a force towards areas of high resource concentration both within a spatial and temporal neighborhood. Note that the limits of the second integral ensure that agents who predict $\tau$ ahead of time are attracted only to the resources in regions to which they can travel within $\tau$. These regions constitute a ``light cone,''  the slope and height of which defined by $c$ and $\tau$. We refer to $\tau$ as the \emph{predictivity} of the agents. 

The assumption of constant velocity  is not a strong one, since an agent can still effectively slow down by going back and forth. However they cannot exceed a maximum speed $c$. We have required that all agents move at a constant velocity so we can fairly compare them to gradient climbing, non-predictive agents. We imagine that the agents discern which direction maximizes their resource consumption, and then move in that direction as quickly as they can. Assigning predictive and non-predictive agents the same maximum velocity also allows us to discern the effects of prediction making. Note that the interaction between agents is indirect: agents are influenced by other agents only through estimating their influence on the resource field. In this sense, the resource field can be thought to mediate interactions between agents in a non-local fashion.

\subsection*{Iterative Solution Scheme}

As our equations are analytically intractable, we solve them numerically through an agent-based simulation. This simulation should be thought as a procedure carried out by individual agents, in order to decide where to move. In the simulations, each agent is represented by a point in a unit square with periodic boundary conditions. Our default time step is $dt=0.001$, and the simulation lasts for $t=1$. In each simulation, the initial distribution of the agents is uniformly random. See the section ``Technical Details'' for additional information about the simulator and numerical methods. Our source code can be found online \cite{GithubCode}.

We solve (\ref{resourceEvolution}) and (\ref{MEquation}) on behalf of the agents, using an iterative scheme. To initialize the process, agents first consider what would happen if  others simply ascend gradients of the resource field while consuming it as they move. This yields a zeroth order estimate for the trajectory $r^{(1)}(x,t)$ of the agents and a zeroth order estimate of what the resource evolution is, $\phi^{(1)}(x,t)$. The agents then set to carry out a more accurate calculation. They ``reset'' all the agents and the resource field to their original state, and then calculate the trajectory $r^{(2)}(x,t)$ of agents according to Eq \ref{MEquation} using $\phi^{(1)}(x,t)$ as an input. As agents move, they consume the resource, the evolution of which constitutes the next order estimation, $\phi^{(2)}(x,t)$, which will be used to estimate $r^{(3)}(x,t)$. They continue this procedure, obtaining at the end of each iteration a new guess for the time evolution of the resource distribution. Throughout, we refer to the index $i$ of $\phi^{(i)}$ and $r^{(i)}$ as the ``(solution) iteration'' coordinate and  can be thought as an internal degree of freedom of agents. We call the number of times this iterative process is carried out the number of \emph{solution iterations} the agents use.

Note that by using the actual simulated data of the agents' motion and the resulting resource distribution, we are assuming that the predictive agents are completely aware of what is happening across all  space. Furthermore, in simulations where we explore the dynamics of a mixture of predicting and non-predicting (i.e. gradient ascending) agents, we assume that the predictive agents are aware of who is predictive and who is not.

We study and report (a) the convergence of the predictive process of the agents, and (b) how well the iterative solution performs, defined in terms of how much resource the agents consume when traversing the solutions they obtain. 

If the solutions $r^{(i)}(t)$ converge with $i$ (or fluctuate with small amplitude), then this means that an agent can exactly (or approximately) predict the future. If the iteration diverges or fluctuates with a large amplitude, the agent cannot make a self-consistent prediction. Convergence and self-consistency are proxies for measuring how well the agents are able to predict the future. 

In our simulations, agents' solutions never converge \emph{exactly}. Nevertheless, if an agent finds a path that is ``approximately correct'', then we can say it is making an accurate prediction. Since successive iterations of path finding can be interpreted as a process of guessing and checking what one's path should be, self-consistency corresponds to agents being able to make more accurate predictions. In other words, a convergent trajectory implies that agents are fulfilling the model and have a correct prediction of the future.

\subsection*{Analysis}
We performed a large number of simulations to determine if and how well predictive solutions converge and how predictive agents perform when competing against non-predictive ($\tau=0$) agents. In every simulation, there were $N=5000$ total agents. The consumption rate of the agents was taken to be $\gamma=1$, and the diffusion rate was $D=0.01$.

\begin{figure}[!h]
\includegraphics[width=\linewidth]{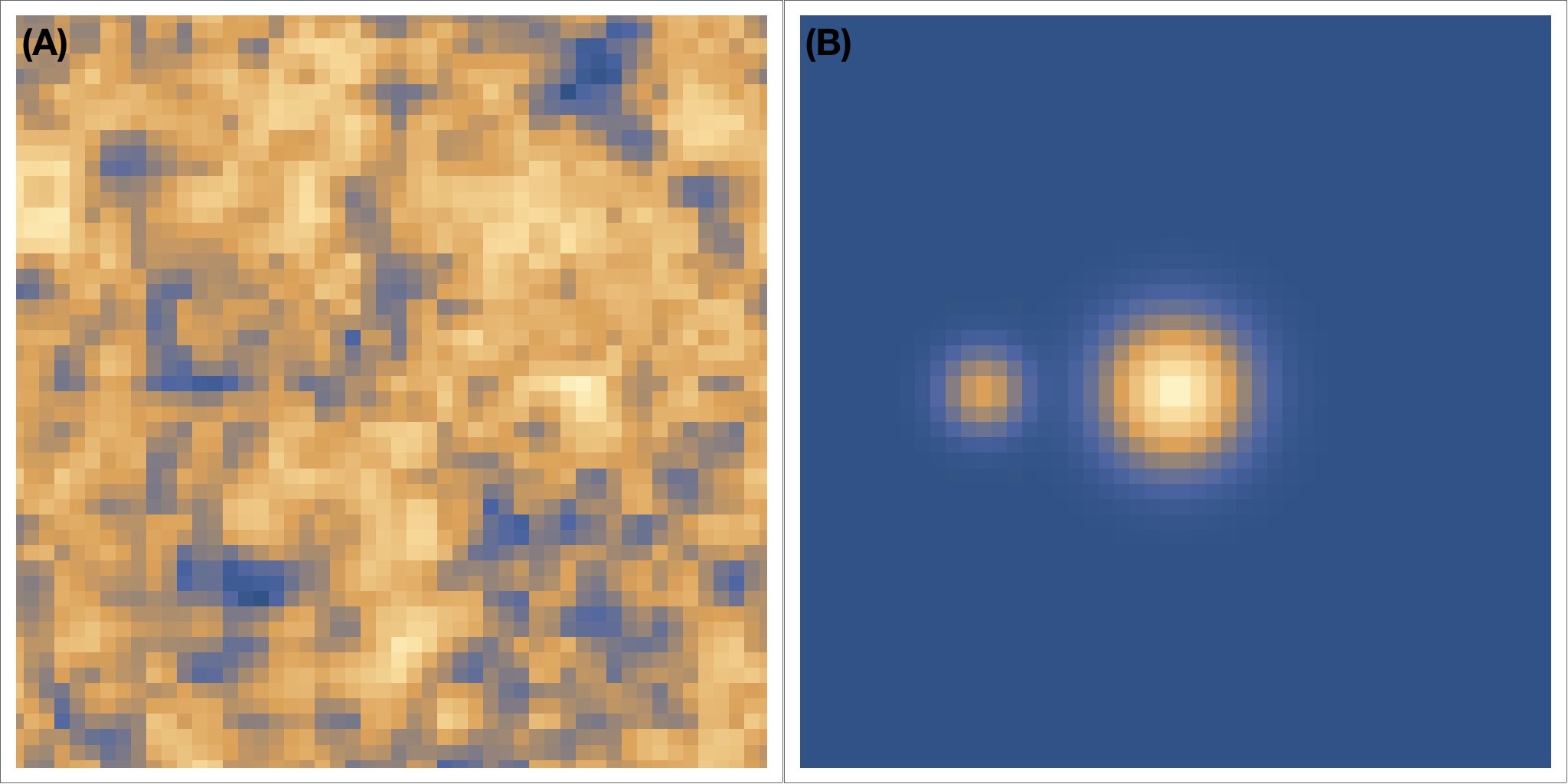}
\caption{{\bf Examples of initial resource.} (Color Online) {\bf Fig 1 A} (Left) Randomly distributed (Perlin Noise) initial resource distribution. {\bf Fig 1 B} (Right) The two peaks initial resource distribution.}
\label{Resources}
\end{figure}

We investigated two initial conditions for the resource distribution. The first initial condition consists of two resource peaks of different magnitude (a larger Gaussian peak in the center and a smaller Gaussian peak to its left). The second is a random distribution (Perlin noise) of resources. Fig \ref{Resources} shows these initial conditions.

We also ran a set of simulations where predictive agents compete against non-predictive agents, which travel at the same velocity, $c$. The non-predictive agents simply ascend the resource gradient, i.e. moving in the direction of $\nabla \phi(\vec{r},t)$. It is straightforward to show  (cf. Appendix) that in the limit $\tau\to0$, predictive agents behave exactly like gradient-ascending, non-predictive agents.

Instead of recording the absolute amount of resources consumed, we obtain a ``consumption factor'' $C_p$ and $C_g$ for the predictive and gradient agents respectively, which we define as the fraction of the total resources consumed by a population (of predictive agents or gradient agents) divided by what fraction of the total agents the population is. To be specific, the consumption factors (CF) $C_p$, $C_g$ of the predictive and gradient ascending (non-predictive) agents are,
\begin{align}
C_{p,g}=\frac{R_{p,g}/R_{tot}}{N_{p,g}/N_{tot}}
\label{ConsumptionFactor}
\end{align}
where $R_p$ and $R_g$ are the amounts of resources consumed, $N_p$ and $N_g$ are the numbers of predictive and gradient agents, and $R_{tot}$ is the total amount of resources available for consumption. The consumption factor allows us to easily compare how well the agents do relative to each other, and simply reduces to the fraction of the total resources consumed when one of $N_p$ or $N_g$ is zero.

We also tracked the difference between agents' paths at different iterations. This difference, for the $k^{th}$ agent between the $m^{th}$ and $n^{th}$ iteration is given by 

\begin{align}
l_k^{(m,n)}=\bigg(\frac{1}{T}\int\limits_{0}^{T} \mathbf{d}t(\vec{r}_m^{(k)}(t)-\vec{r}_n^{(k)}(t))^2\bigg)^{1/2}
\label{L2Difference}
\end{align}

We then average of this quantity over all agents to determine the average $L^2$ path difference between these iterations. If this quantity is small, then agents are, on average, following the same path both iterations. This $L^2$ path difference quantifies how much on average the agents' solutions are changing between solution iterations. We will use it to measure the average difference between consecutive solution iterations, i.e. $L^2 = \frac{1}{S\times N}\sum_{m=1}^{S} \sum_{k=1}^N l_k^{(m,m+1)}$. $L$ measures how well the solutions of agents converge. In all plots the sample standard deviations are displayed as shaded regions around the curves in each figure.

\section*{Results}
\subsection*{Predictive Population}
\begin{figure}[!h]
\includegraphics[width=\linewidth]{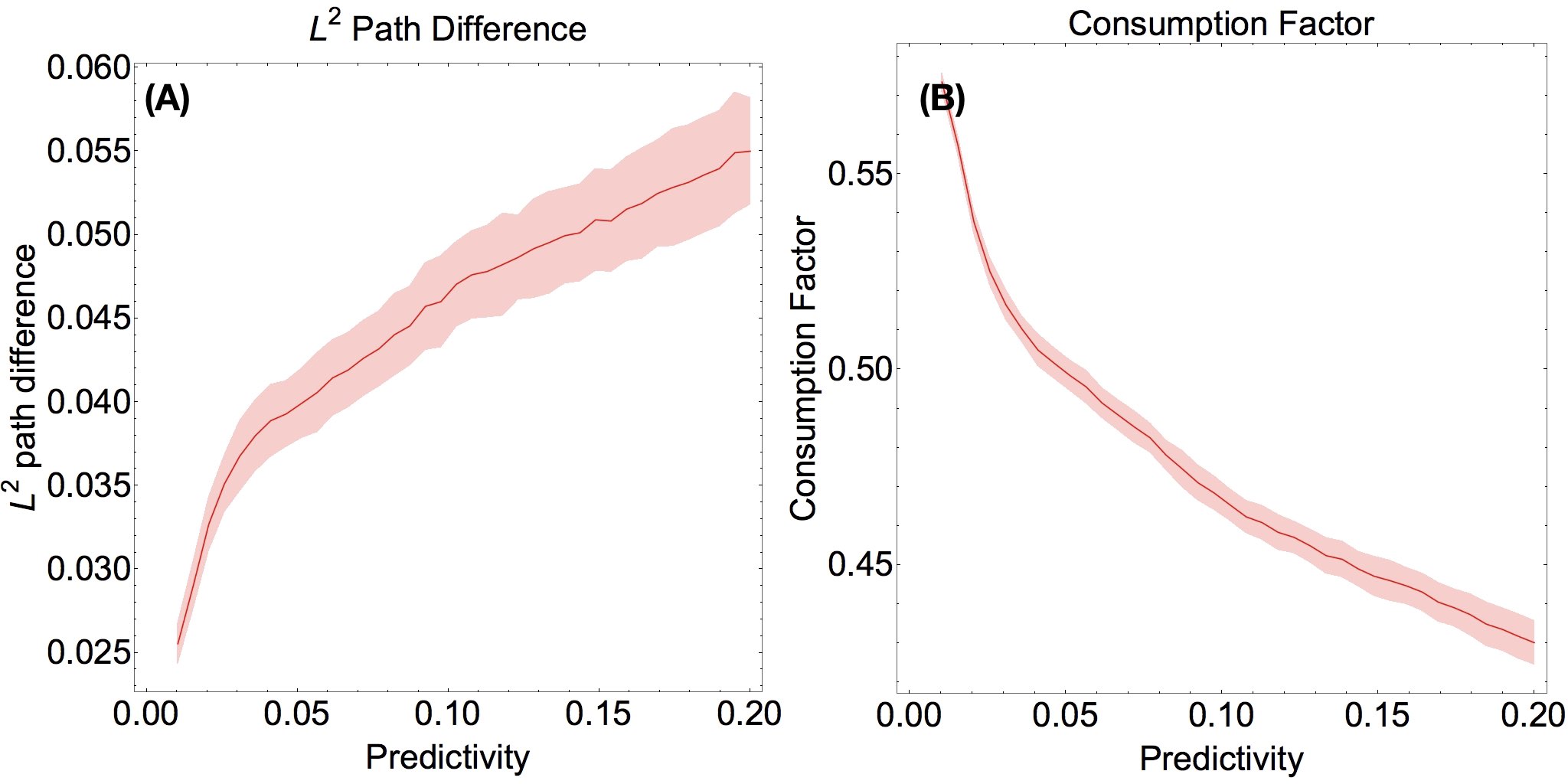}
\caption{{\bf Predictive Agents.} (Color Online) {\bf Fig 2 A} (Left) The $L^2$ path difference vs. predictivity. {\bf Fig 2 B} (Right) The difference in consumption factor vs. agent predictivity. Both plots averaged over 25 runs on a noise resource.}
\label{Predictors}
\end{figure}

We first address how well a population consisting solely of predictive agents performs. We found, counterintuitively, that predictivity is detrimental to consumption success. Specifically, we observed that as the predictivity increases the consumption factor decreases and fluctuates with larger amplitude (Fig \ref{Predictors} A). We measured the average $L^2$ path difference between consecutive solution iterations, and found it to increase with predictivity. It is also clear from Fig \ref{Predictors} B that the consumption factor decreases sharply with increasing predictivity. This indicates that agents are not successfully converging to solutions, except perhaps for very small values of predictivity. More importantly, it is also clear that the agents' consumption factor dramatically decreases with predictivity.

\subsection*{Mixed Population}
The second question we addressed is how well predictive agents perform relative to non-predictive agents when they are mixed together. We ran simulations where the total number of agents are kept constant while the fraction of predictive agents is varied, and measured the solution convergence and resource acquisition success. For each measured quantity we collected, we plot the average value and the sample standard deviation. 

While running simulations of agents on the two peaks resource, we noticed that the performance of the agents depends strongly on the initial positions of the agents, especially when there are very few of them. This is because agents that are close to the peaks will have higher likelihood of consuming a larger amount of resource compared to agents far away from the peaks, regardless of the predictivity of the agents. Thus, to decouple the effects of predictivity from the initial condition, we measured the average improvement in consumption factor due to predictivity, i.e. the consumption factor after a large number of iterations (averaged over a large number of time steps) minus the consumption factor at the first iteration. Since all agents behave as gradient agents in the first iteration, this quantity measures how well predictive agents do compared to how well they would have done had they all been gradient agents.

\begin{figure}[!h]
\includegraphics[width=\linewidth]{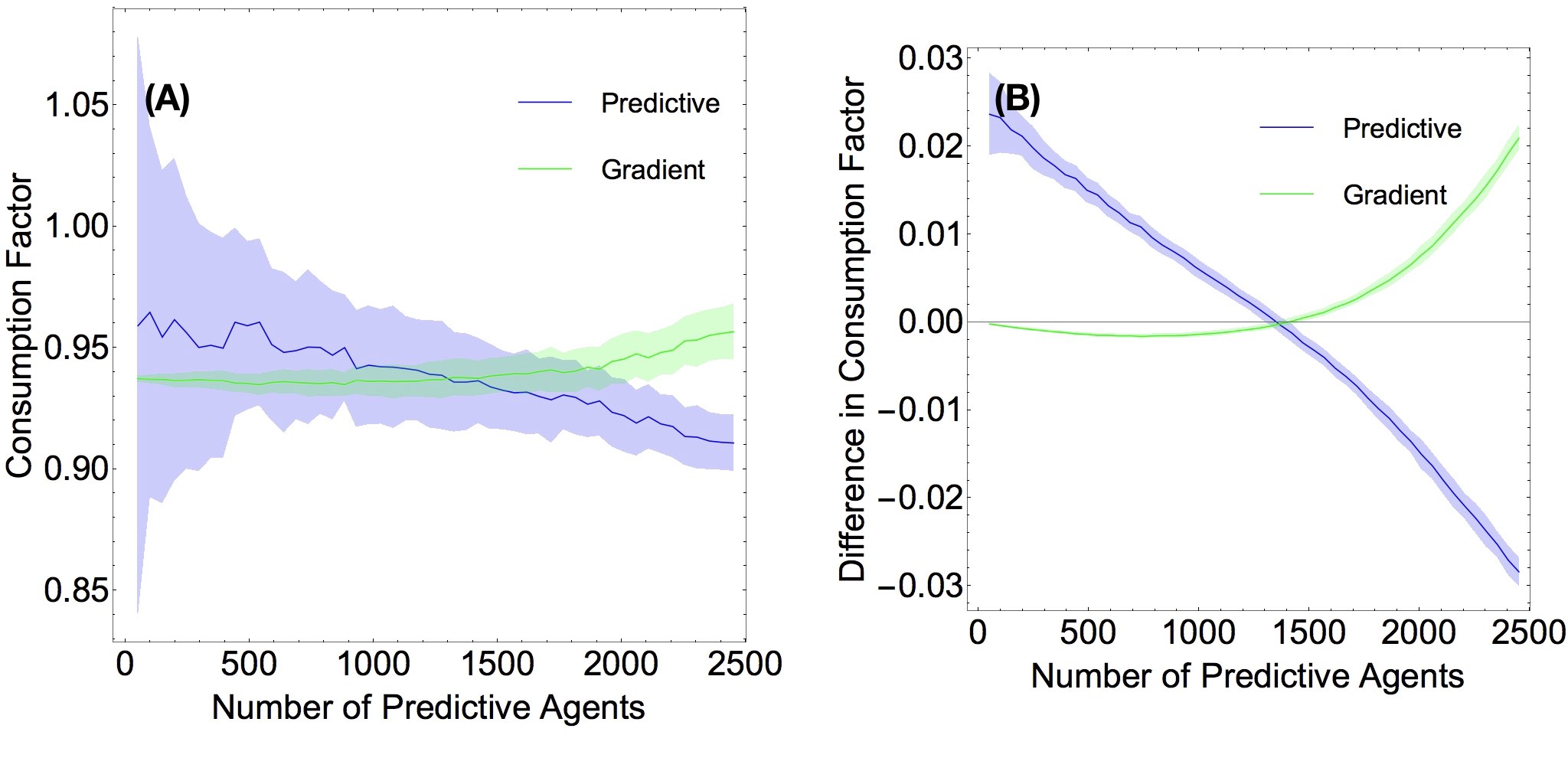}
\caption{ {\bf Comparison of consumption factor and difference in consumption factor.} (Color Online) {\bf Fig 3 A} (Left) Consumption factor vs. number of predictive agents {\bf Fig 3 B} (Right) Difference in consumption factor (how much predictivity improves consumption) versus number of predictive agents. Data was averaged over 200 runs on a two peaks resource, the predictivity of the (predictive) agents was $\tau=0.1$.}
\label{Comparison}
\end{figure}

The difference between these two measures can be seen in Fig \ref{Comparison}. In Fig \ref{Comparison} A we show a plot of consumption factor vs. number of predictive agents. While it is clear that predictive agents do better in small numbers, it is also clear from the standard deviation window that initial conditions play a significant role in how well agents perform. In contrast, when we plot the improvement of consumption factor due to predicting ahead, as in Fig \ref{Comparison} B, the behavior has a much smaller variance.

\begin{figure}[!h]
\includegraphics[width=\linewidth]{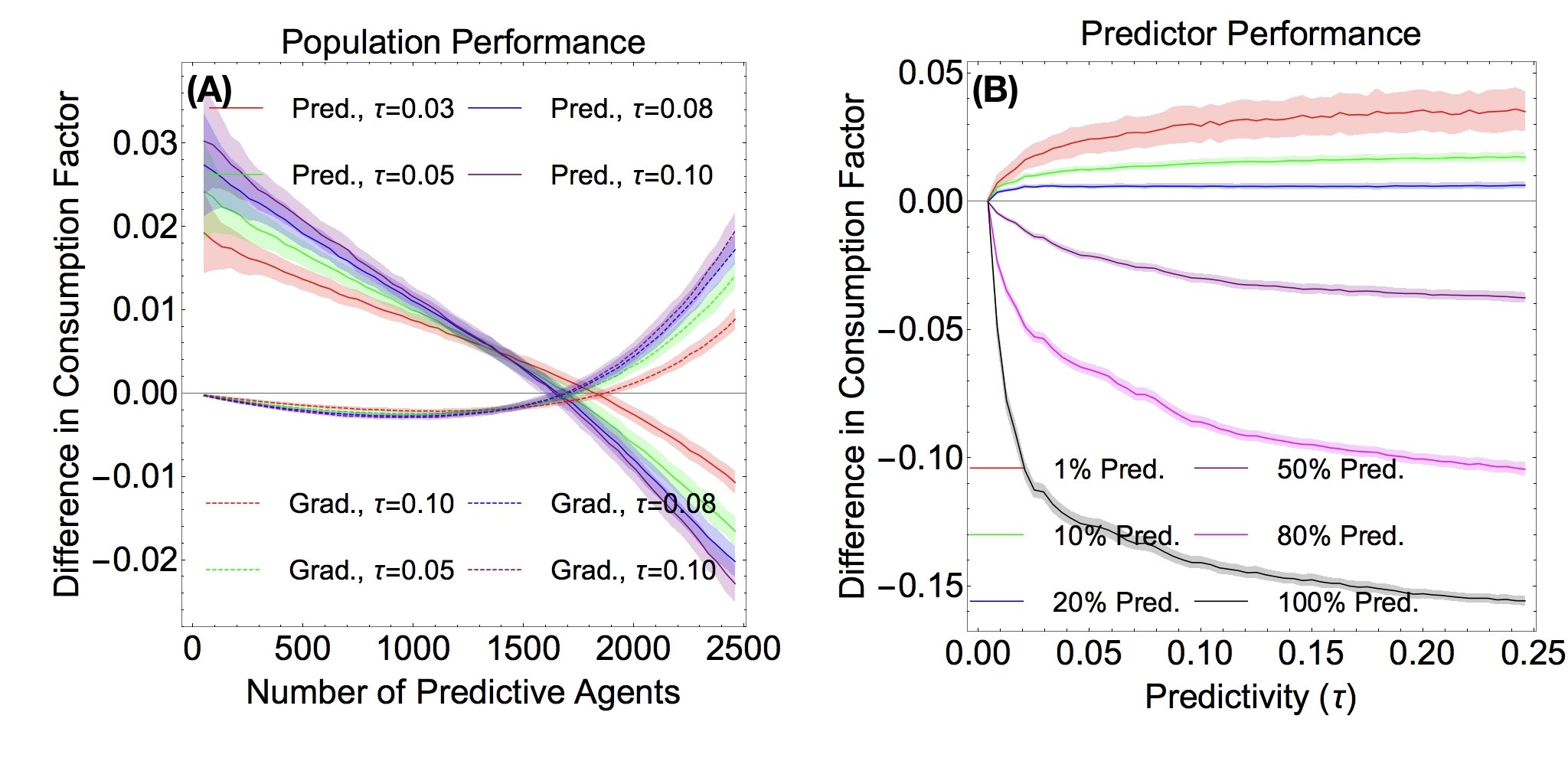}
\caption{{\bf Difference in Consumption Rates.} (Color Online) {\bf Fig 4 A} (Left) Predictive agents consume more resources per capita than gradient agents when there are a small number of them. When there are very few predictive agents, increased predictivity leads to increased performance. {\bf Fig 4 B} (Right) When there are few predictors, increased predictivity is beneficial. When a larger fraction of the agents are predictors, increasing predictivity hurts performance quickly. For both plots, data was averaged over 200 runs on a two-peaks resource.}
\label{Fig2}
\end{figure}

We observed that in the limit where there are few predictive agents, the gradient agents are outperformed by the predictors, as seen in Fig \ref{Fig2} A. However, as the number of predictive agents increases, their advantage is mitigated, and they start performing more poorly then the gradient agents. There are three noteworthy trends here: first, when there are very few predictors (the left-most region of the graph in Fig \ref{Fig2} A), increasing the predictivity increases performance. Second, the larger the predictivity, the fewer predictive agents suffices for the predictive population to perform worse than the gradient population. For example, the crossover for $\tau=0.03$ agents is at around 1850 predictive agents, while the crossover for $\tau=0.1$ is around 1650. Thirdly, there seems to be a point at around 1400 predictive agents where the agents of all predictivities do equally well, as shown by all four lines intersecting with one another. These plots show data with the two-peaks initial condition. The initial condition with Perlin noise exhibits the same qualitative behavior.

We also determined the consumption factor of the predictive agents as a function of predictivity for different percent-compositions of predictors (Fig \ref{Fig2} B). In this case, we observed that predictivity is beneficial only for small percentages of predictive agents. For example, the $1\%$ predictive population benefits from all the values of predictivity in the range we plotted, while the $20\%$ predictive population seems to be helped by having predictivity of up to $\tau=0.05$ or so.

\begin{figure}[!h]
\includegraphics[width=\linewidth]{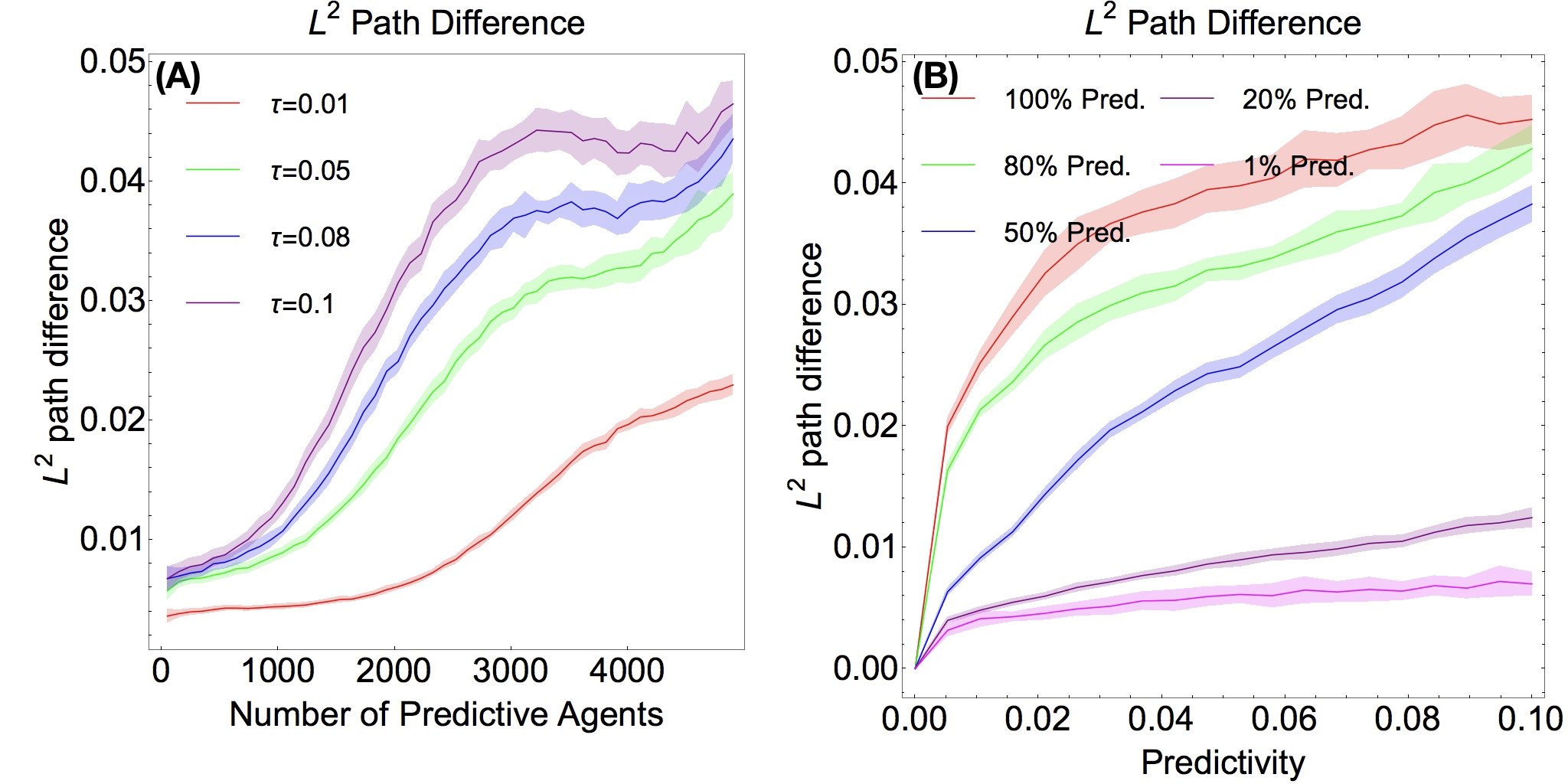}
\caption{{\bf $L^2$ Path Differences.} (Color Online) {\bf Fig 5 A} (Left) The $L^2$ path difference of different predictivity agents, varying what fraction of the total population is predictive. Note that in contrast to Fig \ref{ConsumptionFactor}, we have plotted the data for fractions from 0 to 1 instead of from 0 to 0.5 (there are 5000 agents total). {\bf Fig 5 B} (Right) The $L^2$ path difference for different fractions of predictive agents, varying predictivity. For both plots, data was averaged over 25 runs on a random resource distribution (generated as Perlin Noise).}
\label{L2Plots}
\end{figure}

To determine how much agents' solutions changed iteration to iteration, we analyzed the average $L^2$ difference between agents' paths between iterations. We see in Fig \ref{L2Plots} A that the $L^2$ path difference increases with predictivity and with the percent of predictive agents in the population. In Fig \ref{L2Plots} B, we report the $L^2$ path difference for different population percentages as predictivity is varied. Increasing predictivity still seems to increase the path difference, albeit not too strongly when predictors are in the vast minority. However, it is clear that populations with a large fraction of predictors have larger path differences than populations with very small fraction of predictors.

\subsection*{Effects of Noise}
Noise can drastically affect the collective behavior of many-body systems. To determine if this is the case for our model, we introduced Gaussian white noise to the motion of the agents. More specifically, after observing that predictors often iterated back and forth between the same resource peaks crowded by other predictors, at the same time, we asked whether a weak noise would break this synchrony. We chose a random velocity perturbation of the form $\nu \sqrt{2 D_t \epsilon} \times \vec{v}_r$, where $\nu$ is a random value drawn from a normal distribution, $\vec{v}_r$ is a random unit vector, $D_t = \frac{T}{6 \eta \pi R}$, and we have chosen $\eta=1.308\times10^{-3}$ (the viscosity) and $R=0.05$. To observe what effects random perturbations have on the predictors, we ran simulations similar to those in the preceding section, pitting predictive and non-predictive agents against each other, this time with finite noise.

\begin{figure}[!h]
\includegraphics[width=\linewidth]{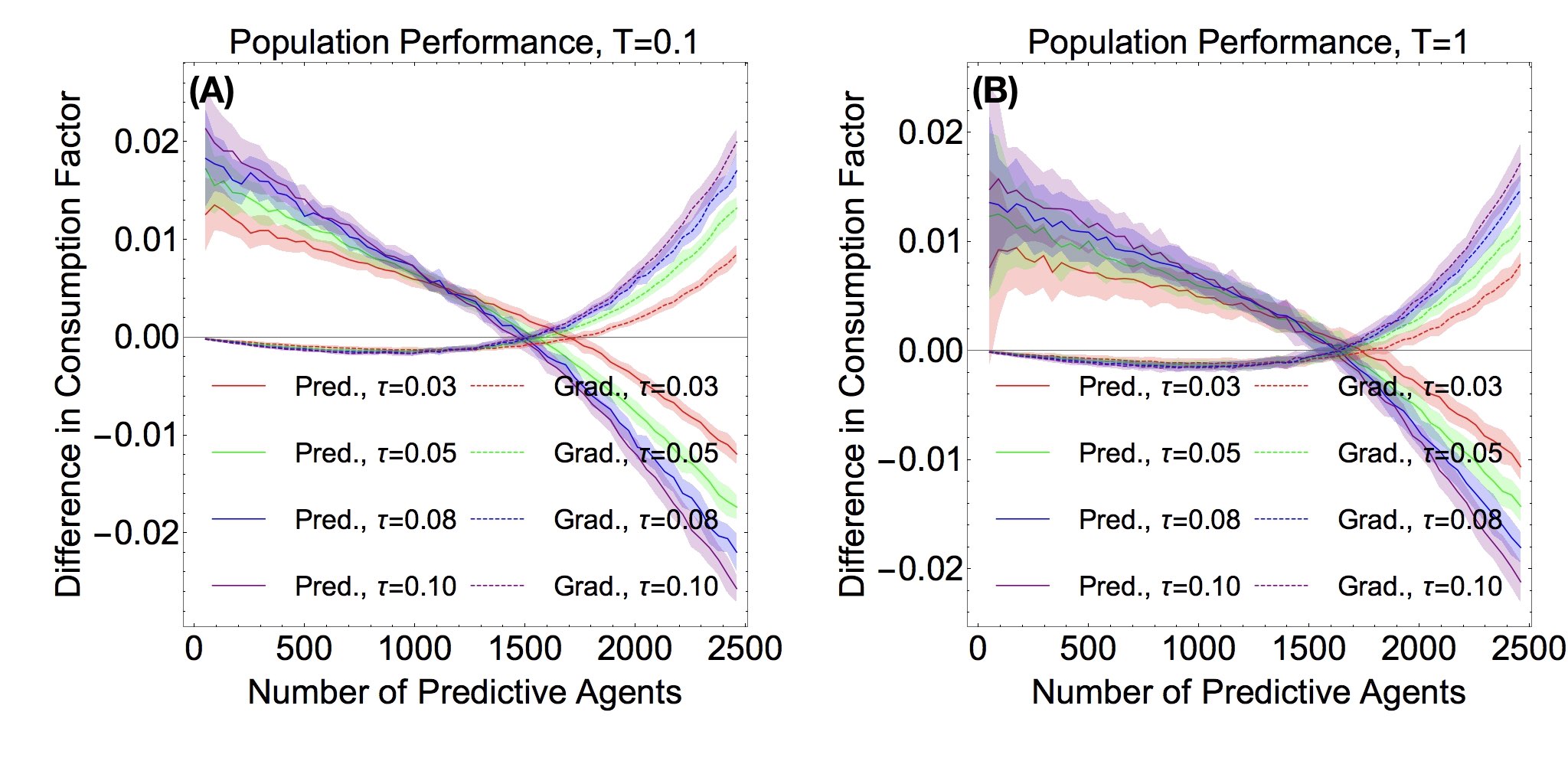}
\caption{{\bf Effects of Noise I.} (Color Online) {\bf Fig 6 A} (Left) Predictors vs. gradient agents (non-predictors) when the noise is $T=0.1$. {\bf Fig 6 B} (Right) Predictors vs. gradient agents (non-predictors) when the noise is $T=1$. For both plots, data was averaged over 50 runs on a two peaks resource.}
\label{Temperature}
\end{figure}

In Fig \ref{Temperature}, we plot the predictor vs. non-predictor behavior for two different levels of noise, $T=0.1$ and $T=1$. We can see that the general behavior of the performance curves is similar to each other and to that which we observe in the $T=0$ case, which we studied in the last section. However, the predictive agents do noticeably worse when there is a non-zero noise, as compared to Fig \ref{Fig2} A, while gradient agents do slightly better than they do at zero noise when there are few predictive agents, and do just as well as before when predictive agents make up a larger fraction of the system. Thus, we find that noise does not correct the ``synchronized poor choices'' of predictors.

\begin{figure}[!h]
\includegraphics[width=\linewidth]{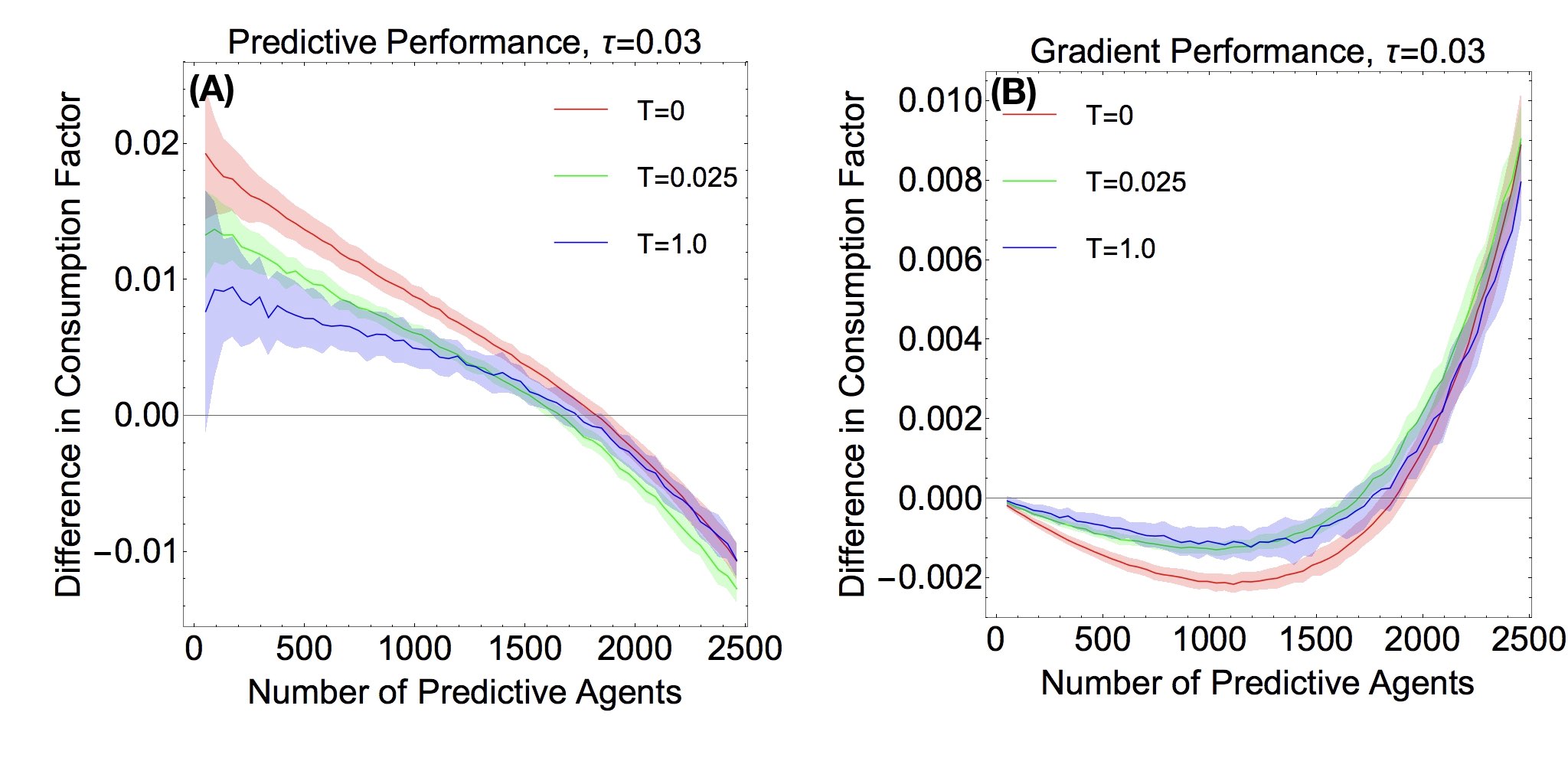}
\caption{{\bf Effects of Noise II.} (Color Online) {\bf Fig 7 A} (Left) Performance of predictors with $\tau=0.03$ at various noise levels. {\bf Fig 7 B} (Right) Performance of non-predictors (competing agains $\tau=0.03$ predictors) at various levels of noise. For both plots, data was averaged over 50 runs on a two peaks resource.}
\label{Temperature2}
\end{figure}

In Fig \ref{Temperature2}, we plot predictor (panel A) and gradient (panel B) performance curves for fixed predictivity, $\tau=0.03$, and various levels of noise. From Fig \ref{Temperature2} A, we can see that the consumption factor decreases monotonically with noise, when there are few predictors. When there are more predictive agents, as seen on the right side of the figure, the curves seem to collapse on one another, indicating that noise does not make a large difference. Note that even a small non-zero noise has a large difference on performance; the difference in performance between $T=0$ and $T=0.025$ is about the same as the difference in performance between $T=0.025$ and $T=1$. In Fig \ref{Temperature2} B, we see the corresponding gradient agent performance curves. We can see that for the fairly small values of noise that we used, the gradient agents have similar curves. We have data for several other noise levels (but did not plot them here, to make the figures more legible), and these curves also show the same behavior, with predictor performance decreasing with increasing noise strength, and gradient performance remaining at a fairly constant level (though slightly above the zero noise level).

\section*{Discussion}

Our results suggest that predictivity, as defined by our model, does not yield success when there are many predictors. Predictive agents do well only when they are competing against a much larger number of non-predictive agents. This is because when there are few predictive agents, they do not affect the system regardless of what actions they take. Thus, they are able to predict the future relatively self-consistently and reliably, to pick the best option for their motion without changing the future they are trying to predict. On the other hand, when there are very many predictive agents, their actions affect the system strongly enough to invalidate their guess at the resource.

Our study of simulation videos of many-body trajectories corresponding to consecutive computation iterations reveals the mechanism behind this prediction driven instability: to maximize resource consumption, agents must move along trajectories that are not only high in resources, but low in population. Thus, it can be advantageous to move towards lower, ``next best'' peaks if these should end up attracting fewer agents. When the density of predictors is low, they can easily see ahead which lesser peaks will be unpopular, directly move there, and out-compete every other agent who moves to the same highest peak.

However when the predictors are densely packed, all predictors close to one another will pick the same ``next-best'' peaks which they anticipate will be least popular. Since this anticipation now increases the peak's future popularity, in the next iteration the predictors will target yet another peak, or return to an earlier choice. 

We have uploaded as supplementary material several videos of the system running. Solution iterations follow one after another, with the positions and resource resetting and then running again. The resource is normalized at each time step, with yellow denoting areas of high resource concentration, and blue areas of low resource concentration. The resource is normalized at each frame - since the resource decays exponentially, this allows us to see the relative abundance of resources for the entire run of the simulation. We denote predictors with purple points and gradient (non-predictors) agents as green points.

We have two videos of mixed populations, S1 and S2 videos, which show mixed populations (with 500 predictors with $\tau=0.05$ and 4500 gradient agents) competing on the two different resources.  While the overall dynamic behavior  is very complex, one can observe that as the predictive iteration proceeds further, a predictor is less likely to occupy the same position as a gradient agent. This indicates that predictive agents have a stronger preference to be at a distance from other agents than gradient agents do. Also note that as iterations proceed, the motion of the resources stay unchanged although the exact positions of the agents are different. This is because of the stability provided by the gradient agents: Since they consume most of the resources, when there are larger number of them, $\phi(x,t)$ becomes more deterministic, and leads the predictive solutions to converge better. While the difference that predictivity makes is subtle, the predictors get a consumption factor improvement of 0.51\% and 0.98\% on the noise and two peaks resource distribution respectively, over the eight solution iterations, and the consumption factor of the gradient agents decreased respectively by 0.028\% and 0.094\%.

We also have two videos of purely predictive agents, S3 and S4 videos. The cause for the poor performance of the predictors is clear: During the first iteration, the predictors behave like gradient agents. As iterations proceed, the predictors cluster around where they think the largest resource peaks will be, and not around lesser peaks. However in the next iteration, these lesser peaks draw the predictors, since agents prefer to move towards peaks that were less popular in their previous iteration. This creates even stronger peaks in other places. In the end, predictors form filaments and clusters that collectively move towards new ``wrong'' places at every iteration. In other words, the agents' predictive solutions are unstable, with small errors tending to amplify and causing their predictions to be worse.

While our simulated model is one among infinitely many very specific choices of models one could devise, we anticipate that the qualitative behavior we have observed will hold for all strongly coupled predictive systems. Specifically, we anticipate that having many predictive agents competing with each other will make it much harder for any of them to predict the future, and that competing against a large number of non-predictive agents will allow predictive agents to accurately predict the future without spoiling their own predictions too much.

Our work is a successor to works discussing unpredictability \cite{moore1990unpredictability} and uncomputability \cite{cubitt2015undecidability,lloyd2016uncomputability,lloyd2013uncomputability,lloyd1993quantum} in physics, and the effects of self-coupling \cite{bennett1995logical}. There are systems whose dynamics cannot be characterized, because doing so would be equivalent to solving the halting problem for a Turing machine, a task that is undecidable. Deciding whether certain quantum spin systems on lattices are gapped or gap-less has been shown to be one such problem \cite{cubitt2015undecidability,lloyd2016uncomputability}, as has the long term behavior of some simple physical models \cite{moore1990unpredictability}. Reference \cite{lloyd2013uncomputability} discusses the fact that some physical systems, such as the aforementioned spin systems, are capable of universal computation, which implies that some quantities associated with the system cannot be known. \cite{lloyd2013uncomputability} gives examples of simple quantum systems where deciding if the energy spectrum is discrete or continuous is uncomputable, and suggests that since algorithmic information content is uncomputable, finding the most concise forms of physical laws may not be possible.

Our agents are running into a fundamental computational problem when they attempt to find solutions to the governing equations (Eqs \ref{resourceEvolution}, \ref{MEquation}). A local differential equation (assuming it is well posed and that there are not to many singularities) can be solved, given a starting point, by finite-difference-type methods. To solve the equation, one takes small local steps, calculating a sequence of functions (each a time slice of the solution) where each function depends on spatio-temporal variables, and the previous function. In other words, the solution can be approximated by a primitive recursive function, i.e. $\phi(k,x_1,...,x_n)$ is defined by $\phi(0,x_1,...,x_n) = \psi(x_1,...,x_n)$ and $\phi(k+1,x_1,...x_n)=\mu(k,\phi(k,x_1,...,x_n),x_1,...,x_n)$ for some functions $\mu$ and $\psi$ \cite{godel1992formally}. Fixed point theory tells us that the error between our functions $\phi(k,...)$ and the true solution is bounded and a decreasing function of the step size. Thus, one can be guaranteed to  solve a differential equation of this kind, within a given error tolerance, in a finite number of steps. 

This is not the case with our system. Barring carefully constructed ``mild'' special cases (such as ones where the agents start very far apart, or they have very small predictivies), our equations do not admit solutions that can be obtained recursively. For such ``mild'' cases that do yield convergent solutions, the only parts of the future an agent has to take into account are parts that looks very similar to the current state of the system, so that our equations effectively become local. However, when the predictivity of the agents start to play an important role, it is not possible to propagate a non-local system forward in time as one usually does with local systems. Instead, one must substitute an entire trajectory and verify if it satisfies the equation.

Unfortunately, this is not a very effective solution method, especially when the non-local system happens to admit chaotic solutions. In this case, a small change between two guessed solutions can lead to vastly different next guesses. What is worse, even if one finds an accurate, self-consistent solution at some level of discretization, this need not be a good solution at a different level of discretization. 

The crucial question then, is, given an initial guess at the system, will our iterative procedure terminate? We have no way to know beforehand if the procedure will terminate, which, loosely speaking, is similar to the halting problem, i.e. it is not possible to have an algorithm that will return whether an input program, together with its initial state, will halt or not \cite{turing1937computable}.

We cannot claim that the difficulties surrounding our chaotic non-local iteration and the limitations of the halting problem goes beyond a resemblance. First, we do not have any proof that it is impossible to solve our system in a finite amount of time. One way to do this might be to showing that the chaotic non-local system at hand is capable of universal computation. Secondly, we do not know if our solution scheme is the most effective one for the problem at hand. It may be that there exists an algorithm that can obtain approximate solutions and converges for all possible initial states and parameter values. Lastly, our problem may simply be ill-posed, in the sense that for certain initial conditions and parameter values, there may not exist any function that satisfies the equations. The existence and uniqueness of differential equations is often proved by iteratively defining sequences of functions with decreasing error and showing that the sequence converges to some function, and that this function fulfills the differential equation. However, this procedure runs into the same issue of not having a recursive way to construct approximations with decreasing error. Accordingly, if the system is uncomputable, it may not be possible to prove that there is a solution.

\section*{Conclusion}

We have constructed a simple, biologically inspired model in which agents attempt to predict the future and act on their predictions to determine the best course of action. We find that convergence of solutions become unlikely with even small amounts of predictivity and that more predictive agents consume less of the resource when they are competing against a large number of other predictive agents. In contrast, when small numbers of predictive agents compete against a large number of gradient agents, they outperform the gradient agents (cf. Fig \ref{Fig2}) and gain more from having greater predictivity. 

While we used a specific model for our simulations, we expect the predictivity driven instability to be a general property for a wider class of systems containing strongly coupled predictors with conflicting interests. When agents in a system base their actions on their prediction of the future (which itself is affected by those actions), the system should exhibit high sensitivity to the initial conditions and the temporal sight of agents.

\section*{Acknowledgments}
Computation was done using the computational resources of the Notre Dame Center for Research Computing.


%
%
%


\section*{Supporting Information}

\noindent { \bf S1 Video. Mixed-Noise } The evolution of a system with 500 predictors and 4500 non-predictors on a noise resource. Eight solution iterations are show, one after another. Green dots represent gradient (non-predictive) agents, purple dots represent predictive agents.

\noindent { \bf S2 Video. Mixed-Two-Peaks } The evolution of a system with 500 predictors and 4500 non-predictors on a two-peaks resource. Eight solution iterations are show, one after another. Green dots represent gradient (non-predictive) agents, purple dots represent predictive agents.

\noindent { \bf S3 Video. Predictors-Noise } A system of 5000 predictors on a noise resource. Eight solution iterations are show, one after another. Green dots represent gradient (non-predictive) agents, purple dots represent predictive agents.

\noindent { \bf S4 Video. Predictors-Two-Peaks } A system of 5000 predictors on a two-peaks resource. Eight solution iterations are show, one after another. Green dots represent gradient (non-predictive) agents, purple dots represent predictive agents.

\section*{Appendix}

\subsection*{Proof that predictive matter reduces to passive matter in the limit $\tau \rightarrow 0$}

We reproduce here, for convenience, Eq \ref{MEquation}:

$$
\vec{M}(\vec{r},t)=\int\limits_{0}^{\tau} d t^\prime\!\!\!\int\limits_{0<|\vec{r}'-\vec{r}|<c\cdot t^\prime} \!\!\!\!\!\!\!\!d^2\!\vec{r}' \cdot \phi(\vec{r}',t+t') \hat{w}(\vec{r}'-\vec{r}) F(R)
$$

and recall that predictive matter moves at a constant velocity $c$ in the direction that $\vec{M}$ points and non-predictive matter moves at a constant velocity $c_p$ in the direction of $\nabla \phi(\vec{r}, t)$. We assume here, as we have throughout the paper, that $c_p=c$. Suppose we take the limit $\tau \rightarrow 0$. We will not explicitly write out $\lim_{\tau \rightarrow 0}$ in the equations below since the actual procedure for determining which direction predictive matter should head involves normalizing $\vec{M}$ before taking the $\tau \rightarrow 0$ limit. We will be using the fact that we only need to keep terms of first order in $\tau$. 

First, with $S_\epsilon(\vec{r})$ being the closed $\epsilon$-ball centered at $\vec{r}$, change position variables in the integration:

$$
\vec{M}(\vec{r},t)=\int\limits_{0}^{\tau} d t^\prime\int\limits_{S_{c\cdot t^\prime}(\vec{r})} d^2\!\vec{r}' \cdot \phi(\vec{r}+\vec{r}',t+t') \hat{w}(\vec{r}') F(R)
$$

where we have shifted our coordinates so $R=R(\vec{r}^\prime,t^\prime)=\sqrt{(r^\prime)^2+\lambda(t^\prime)^2}$. This makes it easy to see that $\phi$ can be expanded as $\phi(\vec{r}+\vec{r}^\prime,t+t^\prime)=\phi(\vec{r},\vec{t})+\vec{r}^\prime \cdot \nabla \phi(\vec{r},t) + t^\prime \frac{\partial \phi}{\partial t} + \mathcal{O}(\tau^2)$. Note that while in general $\vec{r}^\prime$ would not have to depend on $\tau$, our limits of integration, $|\vec{r}|^\prime<c \cdot \tau$, imply that $\mathcal{O}((\vec{r}^\prime)^2)=\mathcal{O}(\tau^2)$. Switching to polar coordinates, using $\hat{w}(\theta) = \hat{x} \sin(\theta) + \hat{y} \cos(\theta)$, and keeping first order terms in $\tau$, our expression for $\vec{M}$ is 

$$
\int\limits_{0}^{\tau} d t^\prime 
\int\limits_{0}^{c \cdot t^\prime} dr 
\int\limits_{0}^{2 \pi} d\theta 
\bigg( \phi(\vec{r},\vec{t})+ r \cdot \hat{w}(\theta) \cdot \nabla \phi(\vec{r},t) + t^\prime \frac{\partial \phi}{\partial t} \bigg) \hat{w}(\theta) F(r)
$$

The first term, $\phi(\vec{r}, t^\prime) F(r) \hat{w}(\theta)$, and the last term, $t^\prime \frac{\partial \phi}{\partial t} \hat{w}(\theta) F(r)$, depend only on theta for each fixed $r$, and integrate to zero in the theta integral. Let us choose our coordinate system so that $\nabla \phi$ points in the $\theta = 0$ direction (the $\hat{y}$ direction). The remaining term does not depend on $t^\prime$, so that integral contributes only a factor of $\tau$, which we shall ignore because we only care about the direction of $\vec{M}$, not its magnitude. 

$$
\vec{M}(\vec{r}, t) \propto
\int\limits_{0}^{c \cdot t^\prime} dr 
\int\limits_{0}^{2 \pi} d\theta 
(F(r) \cdot r \cdot \hat{w}(\theta) \cdot \cos (\theta) \lVert \nabla \phi(\vec{r},t) \rVert)
$$

$$
\vec{M}(\vec{r}, t) \propto \lVert \nabla \phi(\vec{r},t) \rVert
\int\limits_{0}^{c \cdot t^\prime} dr (F(r) \cdot r)
\int\limits_{0}^{2 \pi} d\theta 
(\hat{w}(\theta) \cdot \cos (\theta))
$$

$$
\vec{M}(\vec{r}, t) \propto \pi \lVert \nabla \phi(\vec{r},t) \rVert
\int\limits_{0}^{c \cdot t^\prime} dr (F(r) \cdot r).
$$

Now the remaining integrals are just some real numbers (ignoring the question of convergence, which we will discuss in a moment) and all the vectorial quantities have been found. We then simply have that 

$$
\vec{M}(\vec{r}, t) \propto \lVert \nabla \phi(\vec{r},t) \rVert.
$$

This means predictive agents behave exactly as gradient agents as $\tau \rightarrow 0$. Note that we have not appealed at all to the form of $F$. It is of course possible that $\int\limits_{0}^{c \cdot t^\prime} dr (F(r) \cdot r)$ does not converge. In fact, it would not for our usual force function, $F(r)=r^{-2}$, which is logarithmically divergent. In this case, the correct interpretation is to take the principal value of the integral, i.e. integrating over the annulus $A_{\epsilon, c \cdot t^\prime}(\vec{r})$, normalizing the vector, and only then taking $\epsilon \rightarrow 0$. In this case, the proof is not substantially different. It amounts to changing the lower limit in the previous $r$ integrals for $0$ to $\epsilon$. We get the same result that motion will be in the gradient direction.

The fact that our expression for $\vec{M}$ does not depend at all on the form of $F(r)$ up to order $\tau^2$ tells us that for sufficiently small values of $\tau$, the response of the agents do not depend sensitively on the exact functional form of $F$.

\subsection*{Technical Details}

We give some additional details on the technical aspects of our simulator here for the interested party. The simulator is written in C++ and available for perusal on Github - \url{https://github.com/nrupprecht/Predict}. The most recent version is contained in the folder ``PredictiveSystem,'' while older versions are contained in the folder ``\_PredictiveSystem.''

Resources are consumed in the following manner: for each bin in the (discretized) resource field, a tally is made of all the agents whose positions fall within that bin. The density of agents is calculated, the number of agents in the bin divided by the bin's area. To ensure the performance of the simulation will remain similar with an increasing number of agents, we define a scaled consumption rate to be $\gamma_s = \gamma / N_t$ where $N_t$ is the total number of agents in the simulation and $\gamma$ is the externally set consumption rate parameter. This way, increasing the number of agents has the effect of more closely approximating a continuous distribution of agents, and the characteristic resource consumption time is independent of the total number of agents, only depending on $\gamma$. The amount of resources in the bin is decreased by $\phi_t \rightarrow \phi_t - \epsilon \gamma_s \phi_{t-1} \rho $, and the amount of resources the agents have consumed is increased by $\epsilon \gamma_s \phi_{t-1} N_{bin}$ where $N_{bin}$ is the total number of agents in the bin. With this setup, the consumption rate of the agents and behavior of the resource is independent of the field discretization and the number of agents, and the total reduction in resources over the length of the simulation matches the total amount consumed by the agents. This process is done separately for predictive and gradient agents so we can record the consumption of each population independently. Note that since consumption depends on $\phi_{t-1}$, which is set, and $\phi_t$ is adjusted, no agent consumes any portion of the resource ``before'' any other, so the system is fair.

The program can be compiled by either intel icpc compiler or GNU g++ compiler by changing two lines in the Makefile marked ``CHOSE COMPILER.'' For intel, use ``CC = \$(ICC)'' and ``COMP = \$(ICOMP),'' for GNU, use ``CC = \$(GCC)'' and ``COMP = \$(GCOMP).'' Currently, the intel version is faster than the GNU version.

\end{document}